\PassOptionsToPackage{table}{xcolor}
\documentclass[sigconf]{acmart}
\setcopyright{none}
\renewcommand\footnotetextcopyrightpermission[1]{}
\usepackage{xspace}
\usepackage{amsmath}
\usepackage{graphicx}
\usepackage[colorinlistoftodos]{todonotes}
\usepackage{subcaption}
\usepackage{booktabs}
\usepackage{soul}
\usepackage{tabto}

\title{Security Issues in Language-based Software Ecosystems}
\author{Ruturaj K. Vaidya$^{1}$\ \ \ \ Lorenzo De Carli$^2$\ \ \ \ Drew Davidson$^1$\ \ \ \ Vaibhav Rastogi$^3$}
\author{$^1$University of Kansas\ \ \ \ $^2$Worcester Polytechnic Institute\ \ \ \ $^3$University of Wisconsin, Madison}

\clearpage{}%

\newcommand{\naive}{naive\xspace}
\newcommand{\myparagraph}[1]{\vspace{0.1in}\noindent\textbf{#1}}

\clearpage{}%
\begin{document}
\begin{abstract}
  Language-based ecosystems (LBE), i.e., software ecosystems based on a single programming language, are very common. Examples include the npm ecosystem for JavaScript, and PyPI for Python. These environments encourage code reuse between packages, and incorporate utilities---package managers---for automatically resolving dependencies. However, the same aspects that make these systems popular---ease of publishing code and importing external code---also create novel security issues, which have so far seen little study.

We present an a systematic study of security issues that plague LBEs. These issues are inherent to the ways these ecosystems work and cannot be resolved by fixing software vulnerabilities in either the packages or the utilities, e.g., package manager tools, that build these ecosystems. We systematically characterize recent security attacks from various aspects, including attack strategies, vectors, and goals. Our characterization and in-depth analysis of npm and PyPI ecosystems, which represent the largest LBEs, covering nearly one million packages indicates that these ecosystems make an opportune environment for attackers to incorporate stealthy attacks.
  Overall, we argue that (i) fully automated detection of malicious packages is likely to be unfeasible; however (ii) tools and metrics that help developers assess the risk of including external dependencies would go a long way toward preventing attacks.
\end{abstract}

 \maketitle
\section{Introduction}
A recent report by the software security company Contrast
Security found that 79\% of application code came from third
parties~\cite{contrast}. The use of third-party code has obvious
benefits: it encourages code reuse; it allows expertly-written and
well-vetted codebases to be deployed by more developers; and it
leverages the knowledge of the broader software development
community even for highly-custom projects. However, managing 
third-party components has become increasingly complex. A
complex web of dependencies exists because third party components
internally depend upon one another. Furthermore, these
components update out of step with one another, introducing new
functionality and behavior. 

To ease the complexity and burden of navigating the use of third-party
code, a traditional solution has been to organize third-party
components into \textit{packages}, which provide discrete modules of
functionality.
The
dependencies between packages are listed explicitly as metadata
within the package by third-party developers, and packages are
stored in an online database, or a \textit{package repository}.

Much of the complexity of using packages is delegated to a 
utility program called a \textit{package manager}, which can 
navigate the web of dependencies to find up-to-date versions of 
packages and ensure that package dependencies are provided. 
A key goal of package managers is that they abstract away the 
complexity of integrating third-party functionality into a
software project. However, the key insight of our paper is that
this abstraction introduces the potential for stealthy attacks
that may go undetected for long periods.

In this paper, we specifically study package management for {\em language-based ecosystems} (LBEs),
using the ecosystem of npm for JavaScript/Node.js and PyPI for
Python as case studies. Packages from these ecosystems form the backbone of software
development in those specific languages by hosting third-party code
that is reused in many different software projects. 

There exists some prior work studying software repositories such as mobile app
stores like Google Play and Apple App Store, which serve consumers with
full-fledged applications rather than developers with re-usable code
components, and OS package managers such as RPM and Apt~\cite{viennot2014measurement,chatterjee18-ipv,wermke18,
chakradeo13,cappos_look_2008}. LBEs have 
received much less attention, even though LBEs are inherently different
from other software repositories. We therefore focus our work on attacks that
arise inherently from the way LBEs work. As such, we consider vulnerabilities
in either the packages or the package management system to be outside the scope
of our work. 

Previous work in both the industry and the academia has identified
specific instances of malicious attacks on these package management
ecosystems (e.g.,~\cite{catalin_cimpanu_backdoored_2018,goodin_devs_2017,fallingsnow_i_2018}). Our work is the first to systematically study language-based
ecosystems and presents a holistic
perspective on attacks in these ecosystems by
providing a characterization and taxonomy of attacks, and by
analyzing package repositories based on metrics that relate to
potential for attacks. 

\subsubsection*{Contributions}
The contributions of our paper are as follow:
\begin{itemize}
    \item We introduce a taxonomy of LBE 
	compromises to characterize the landscape of known
	attacks. We then use it to categorize many notable
	examples of such attacks.
    \item We propose metrics for evaluating the risk and the 
	impact of package compromise.
	We believe these metrics serve as a call to action for
	additional work in the domain. 
    \item  We perform case studies to characterize the state
	of two popular package management ecosystems, npm and PyPI. 
	Our broad analysis of these two ecosystems and specific
	case studies serve to demonstrate the use of our 
	metrics and to identify risks and security-relevant factors in
	current ecosystems (such as developer behavior and the
	interconnectedness of packages).
    \item We present concrete proposals for improving the
	security of package managers against a class of attacks unique
	to this domain. Our proposals include best-practices for
	avoiding common mistakes that lead to compromises, and
	enhancements to package manager software so that such
	compromises are easier to avoid. 
\end{itemize}

Our analysis includes npm and PyPI ecosystems, covering about a million packages
overall. We present important insights into how social engineering attacks such
as typo-squatting and import-squatting may actually be flourishing. We find that due
to the nature of these ecosystems, there is no easy solution against such
squatting problems at the ecosystem level. At the same time, through our characterization of past attacks, we find that much of these attacks can be effectively mitigated through simple feedback to developers as they interact with the ecosystems.

The rest of this paper is organized as follows. Section~\ref{sec:background}
provides a broad background on package management ecosystems and describes how
our paper fits in. Section~\ref{sec:attacks} presents a taxonomy of attacks on language
package management ecosystems and characterizes previous attacks based on this
taxonomy. Section~\ref{sec:analysis} further analyzes these ecosystems. In
Section~\ref{sec:recommendations}, we present recommendations based on our
attack characterization and analysis. We then present
related work, relevant discussions, and conclusions in Sections~\ref{sec:related},\ref{sec:discussion}, and \ref{sec:conclusion}, respectively.

\section{Background}\label{sec:background}
Package managers, particularly those that focus on packages from specific
languages, play an essential role in software development. The intention of such
package managers is to provide modules of high-quality code that can be
reused efficiently by other developers. As software complexity and the related
maintenance effort increase, it is all the more desirable to use third-party
code as much as possible and have it automatically managed. Package managers
fill in this desideratum by keeping track of software dependencies (including
dependencies of dependencies) and taking care of compilation and
installation of these dependencies.

Different languages differ significantly in their package management models and
philosophies. Some language ecosystems have a de facto package management
repository. For instance, Python has PyPI, JavaScript/Node.js has npm, Rust has
Crates, and Java (including all JVM languages) has Maven. These languages
then use various package managers such as \texttt{pip} for PyPI, \texttt{npm}
and \texttt{yarn} for npm (i.e., the npm repository), \texttt{cargo} for Crates, and \texttt{mvn} and
\texttt{gradle} for Maven, which pull data from these repositories and provide
other functionality such as handling project builds.  Other languages
such as Go and C++ do not have specific repositories where one may find most
third-party code; instead the communities in these languages prefer to directly
use the disparate project repositories (such as a project-specific GitHub
repository). Moreover, C++ has evolved a large number of different package
managers catering to different philosophies and platforms~\cite{pfultz-upm, antoninj-reddit}.
Finally, the OS package managers (e.g. RPM and Apt) in Unix systems
provide some C/C++ libraries and so play a partial role as package managers for
C/C++. It is impossible to cover the specifics of each language-based package
management in one paper. We therefore focus on a common package management model
where there is one de facto universal repository that hosts most of the
third-party code -- Python, JavaScript, Rust, Java, Ruby, PHP, Perl and many
others follow this model. We study and present our findings on the Python
and JavaScript ecosystems as they are representative of this model and have some
of the largest user base and package repositories.

Apart from package management for software development, there exist other
package managers and repositories. For instance, mobile app stores, in
particular, Google Play and Apple App Store host millions of applications and
are central to the Android and iOS ecosystems. However, the focus of these
repositories is more on the consumer than the developer and hence handle
different challenges and security issues. For instance, while social engineering
attacks would plague both a mobile app store and software package repository
like npm, the specific vectors of attacks will be different: in a mobile app
store, a user may, for example, be enticed into installing a fake version of a
game by deceptive imagery and wording~\cite{fortnite-android} while on a language package
repository, the vector may be typosquatting (Section~\ref{sec:analysis}).
Likewise, traditional OS package managers and repositories like RPM and Apt face
threats that are different from those faced by language package managers. For
instance, OS package managers provide a well-deliberated set of packages with
assigned package maintainers who are often well-known in the community while
language-based ecosystems are often much more relaxed with anyone allowed to
upload packages to the package repositories.

\section{Characterizing Past Attacks}
\label{sec:attacks}
\begin{table*}[!t]
  \caption{Summary of attacks discussed in the paper}
  \label{tab:attacks}
  \vspace{-0.1in}
  \rowcolors{2}{gray!20}{white}
  \begin{tabular}{p{2.25in}p{3.5in}p{0.65in}}
    \toprule
    {\bf Attack} & {\bf Notes} & {\bf Time-to-discovery}\\
    \midrule
    event-stream compromise~\cite{fallingsnow_i_2018} & Attacker turned the copay package into wallet-stealing tool by attacking event-stream, on which copay has an indirect build dependency & 46 days \\
    Go-bindata account takeover~\cite{fox_hijacking_2018} & Github account for developer of popular go-bindata package (unmaintained at the time) re-created after owner deleted it & Same day\\
    mailparser backdoor~\cite{catalin_cimpanu_somebody_2018} & Attacker added a dependency chain to popular but unmaintained mailparser package, terminating with a backdoored package & 20 days\\
    npm ESLint-scope password stealer~\cite{shaun_nichols_now_2018} & Attacker inserted credential-stealing code into ESLint-scope, a dependency of the popular ESLint package which received seldom updates & Same day\\
    conventional-changelog compromise~\cite{tjenkinson_information_2018} & Attacker compromised popular, and actively maintained conventional-changelog package inserting a cryptominer & 1.5 days \\
    npm typosquatting~\cite{npm_authors_`crossenv`_2017} & Attacker uploaded 40 information-stealing packages with names similar to those of popular packages & 12 days\\
    PyPI backdoor~\cite{catalin_cimpanu_backdoored_2018} & Attacker inserted malicious credential-collecting code in the (well-maintained) PyPI ssh-decorate module & 3 days \\
    PyPI typosquatting~\cite{goodin_devs_2017} & Attacker uploaded 10 malicious packages with names similar to those of popular packages & 99 days \\
    \bottomrule
  \end{tabular}
  
\end{table*}
 In this section we discuss the characteristics of eight incidents involving security compromise of package ecosystems. We selected these incidents by reviewing news and reports from the last two years containing descriptions of package manager-related incidents, which resulted in 15 relevant unique incidents. We then discarded 7 incidents because they were out of scope (e.g., attacking software bug in package management software rather than the ecosystem~\cite{max_justicz_remote_2018}), or directed against traditional OS package ecosystems (e.g.,~\cite{tarwirdur_how_2018}). Table~\ref{tab:attacks} summarize these incidents.

\begin{table*}[]
  \caption{Framing of attacks within proposed taxonomies}
  \label{tab:taxonomies}
    \vspace{-0.1in}
  \rowcolors{2}{gray!20}{white}
  \begin{tabular}{p{2.25in}lllll}
    \toprule
    {\bf Attack} & {\bf Type} & {\bf Strategy} & {\bf Vector} & {\bf Victims} & {\bf Goals}\\
    \midrule
    event-stream compromission~\cite{fallingsnow_i_2018}                    & Influencer & social engineering & package code & 2nd-party & crypto theft \\
    Go-bindata account takeover~\cite{fox_hijacking_2018}                   & Direct     & social engineering & N/A & 1st-party & unknown\\
    mailparser backdoor~\cite{catalin_cimpanu_somebody_2018}                & Influencer & credential stealing & package code & 2nd-party & credential theft \\
    npm ESLint-scope password stealer~\cite{shaun_nichols_now_2018}         & Direct     & credential stealing & installation script & 1st-party & credential theft\\
    conventional-changelog compromise~\cite{tjenkinson_information_2018} & Direct     & credential stealing & package code & 1st-party & crypto theft \\
    npm typosquatting~\cite{npm_authors_`crossenv`_2017}                    & Bait       & social engineering & installation script & 1st-party & credential theft \\
    PyPI backdoor~\cite{catalin_cimpanu_backdoored_2018}                    & Direct     & credential stealing & package code & 2nd-party & credential theft\\
    PyPI typosquatting~\cite{goodin_devs_2017}                              & Bait       & social engineering & installation script & 1st-party & dry run\\
    \bottomrule
  \end{tabular}

\end{table*} 
\subsection{Overview}

\subsubsection{Threat Model}

In our work, we focus on attacks that are systemic to the LBEs themselves, i.e., they exploit some properties of the ecosystems such as the \textit{densely connected nature of LBEs and their dependency graphs}, and \textit{low barriers to entry for publishing packages}. Therefore, we consider incidents involving an attacker with any (or all) of the following capabilities:

\begin{enumerate}
\item The attacker has the capability to publish an arbitrary number of new packages.
\item The attacker can compromise existing developer accounts via either social engineering, brute-forcing, misconfiguration in package management infrastructure, or credential reuse\footnote{While not every developer account can be compromised by these means, review of past incidents shows that a non-trivial portion of developer accounts on popular websites is vulnerable~\cite{chalker_public_2018}.}.
\end{enumerate}

In all but one incidents considered, the attacker uses the means above to create and hide malicious code in a package. Code may be hidden in the application itself, or in the installation scripts shipped with the package. We elaborate further on both approaches later in this section.

\subsubsection{Known Attack Instances}

Due to space constraints, we only discuss three representative incidents from our dataset in detail. Readers may consult the respective references for other incidents.

\myparagraph{npm typosquatting~\cite{npm_authors_`crossenv`_2017}.} Between July 19th and 31st 2017, the npm user account ``HackTask'' uploaded malicious packages to npm, the de-facto standard package manager for Node.js. These packages had names similar to existing, benign packages---an attack known as \textit{typosquatting}. For example, one of the malicious packages was named \texttt{crossenv}, similar to the popular utility \texttt{cross-env}. All the malicious packages included an attack payload in one of the installation scripts; the payload was designed to ex-filtrate local environmental variables (which oftentimes on development machines store sensitive authentication tokens) to an attacker-controlled location. The attack lasted 12 days before being discovered, and npm analysts estimate that around 50 developers mistakenly downloaded malicious packages instead of the original ones, and were thus affected by the attack.

\myparagraph{ESLint-scope password stealer~\cite{shaun_nichols_now_2018}.} In the night between July 11th and 12th 2018, an attacker used a compromised developer account to publish a malicious version of the \texttt{eslint-scope} npm package (a submodule of ESLint, a popular JavaScript code analysis toolkit).  The package was altered to download and execute a credential-stealing payload on installation. The payload copies npmjs.org login credentials---stored in the .npmrc file---to an attacker-controlled server. The attack was detected almost immediately, but the ESLint development team estimated that up to 4,500 accounts may have been compromised~\cite{eslint_authors_postmortem_2018}.

\myparagraph{event-stream compromise~\cite{fallingsnow_i_2018}.} This case study consists of a complex, multi-stage attack against the users of the \texttt{copay} npm package. In 2018, an attacker emailed the developer of a popular but unmaintained npm package, \texttt{event-stream}, on which \texttt{copay} has an indirect build dependency. The attacker offered to help with development. After achieving access, on September 9th 2018 the attacker injected a dependency on an external package in \texttt{event-stream}. On October 5th, the application code of the external dependency (\texttt{flatmap-stream}) was padded with a malicious payload. When certain operations are executed specifically during the \texttt{copay} building process, the \texttt{copay} build was altered so that, when installed and used, it ex-filtrated sensitive wallet information to an attacker-controlled server. The attacker went undiscovered for 46 days, and multiple official \texttt{copay} releases (5.0.2 to 5.1.0) were affected~\cite{npm_authors_details_2018}. The amount of cryptocurrency stolen in the attack is unclear.

Review of the three cases above suggests that attacks against package ecosystems can exhibit significant differences in terms of approach and goals. In order to understand these difference and support our analysis, we therefore developed a set of attack taxonomies, which we illustrate below. Table~\ref{tab:taxonomies} frames the attacks under analysis within these taxonomies.

\subsection{Attack types}
\label{sec:AttackTypes}
In this work, we use various specific classes of attacks:

\begin{itemize}
\item \textbf{Bait attack:} an attacker creates an appealing package, i.e. one which is likely to be of interest to the community due to some package features. One example of this category is the typosquatting attacks~\cite{npm_authors_`crossenv`_2017}, where a package is named similarly to another, popular package. 
\item \textbf{Direct attack:} an attacker fraudulently and directly gains access to the target package, and proceeds to inject malicious payload in the code. 
\item \textbf{Influencer attack:} an attacker targets a package by inserting malicious code into another package on which the victim package depends. Attackers can leverage existing dependencies, or inject new ones into the victim package through various means. Both direct and build dependencies have been leveraged in past incidents. %
\end{itemize}

\subsection{Strategy and factors affecting success}

While bait attacks leverage unprivileged social engineering, direct and influencer attacks require gaining some form of access and publishing rights for an existing software package. Attackers may break into existing developers' accounts~\cite{chalker_public_2018} (e.g., by exploiting credential reuse), or simply volunteer to help~\cite{fallingsnow_i_2018}. A different approach entails re-creating a package under the name of a popular one which has been removed~\cite{fox_hijacking_2018}.

All attacks need to deploy malicious code before being detected. Code obfuscation and payload encryption are sometimes used for this purpose~\cite{fallingsnow_i_2018}. For direct and influencer attacks, the attacker may also attempt to select target packages which have gone unmaintained~\cite{catalin_cimpanu_somebody_2018}, in the hope that malicious commits may undergo less scrutiny. We note, however, that actively developed packages have been targeted too~\cite{tjenkinson_information_2018}. While evidence is too limited for quantitative analysis, data in Table~\ref{tab:attacks} suggests that deliberately choosing to target unmaintained packages may prolong the attack.

A related challenge for the attacker is ensuring that attack code is spread widely\footnote{While targeted attacks are in principle possible, we have not observed them in our dataset. We note, however, that such attacks would be more likely to go undetected.}. In a bait attack the attacker attempts to engineer packages that are likely to appeal to the community. The attacks we consider achieve this by typosquatting, however alternative approaches are possible. Examples include creating a package purporting to offer a desirable functionality~\cite{tarwirdur_how_2018}, or artificially inflating package popularity metrics (e.g., by repeatedly downloading it, or sending pull requests that inflate the number of packages depending on it~\cite{gilbertson_im_2018}). In direct and influencer attacks, attackers tend to choose victim packages which are either already popular~\cite{fallingsnow_i_2018}, or are dependencies of other, popular packages~\cite{shaun_nichols_now_2018}.

\subsection{Attack vector}

For attacks that do attain the goal of injecting malicious code, it is useful to consider where the malicious code resides:

\begin{itemize}
\item \textbf{Package code:} attack payload executes when the package is loaded or functions within the package are executed. For indirect attacks, the attack code resides within a dependency which must be imported and used for the attack to work.

\item \textbf{Installation script:} virtually all package managers offer the ability to execute custom scripts during various stages of package installation. Injecting attack code within these scripts ensures that the attack is executed even if the package code itself is never actually used.
\end{itemize}

It should be noted that, regardless of how attack code is executed, it is in principle possible for the payload to persist the attack, e.g. by creating init scripts to re-execute the attack at every boot~\cite{tarwirdur_how_2018} or by adding code \texttt{.bashrc} or \texttt{.bash\_profile}~\cite{mitre-t1156}.

\subsection{Attacker victims and goals}

We define two classes of attack victims:

\begin{itemize}
\item \textbf{First-party victims:} developers taking part in and contributing to the package ecosystem.
\item \textbf{Second-party victims}: users of applications based on compromised packages.
\end{itemize}

While attacks against second-party victims have been rare so far, they outline a concerning scenario. In fact, these attacks may affect a large number of nontechnical users who are unfamiliar with the package ecosystem and the risk of attacks. Next, we look at possible attacker goals:

\begin{itemize}
\item \textbf{Dry-run:} we include in this category attacks with a payload with limited or no malicious effect. Such attacks may have been carried by researchers, or by cybercriminals evaluating their tools.
\item \textbf{Cryptocurrency theft:} this type of attack aims at injecting code which either executes cryptomining, or attempts to steal funds. The former goal in particular is ideally suited to the characteristics of these attacks, which are typically discovered quickly but may reach a large number of users.
\item \textbf{Credential theft:} this type of attacks attempts to identify and appropriate credentials and/or sensitive information of either developers or application users.
\end{itemize}

Our analysis suggests that security risks in LBEs chiefly originate from three factors: the ecosystems' \textit{scale}, their \textit{interconnectedness}, and the prevalence of package \textit{abandonment}.
In the next section, we consider metrics to capture each factor.
We then perform ecosystem-wide analyses for typosquatting attacks. Many of the real-world incidents belong this attack class.
\begin{figure}[t]
  \centering
  \includegraphics[width=0.5\textwidth]{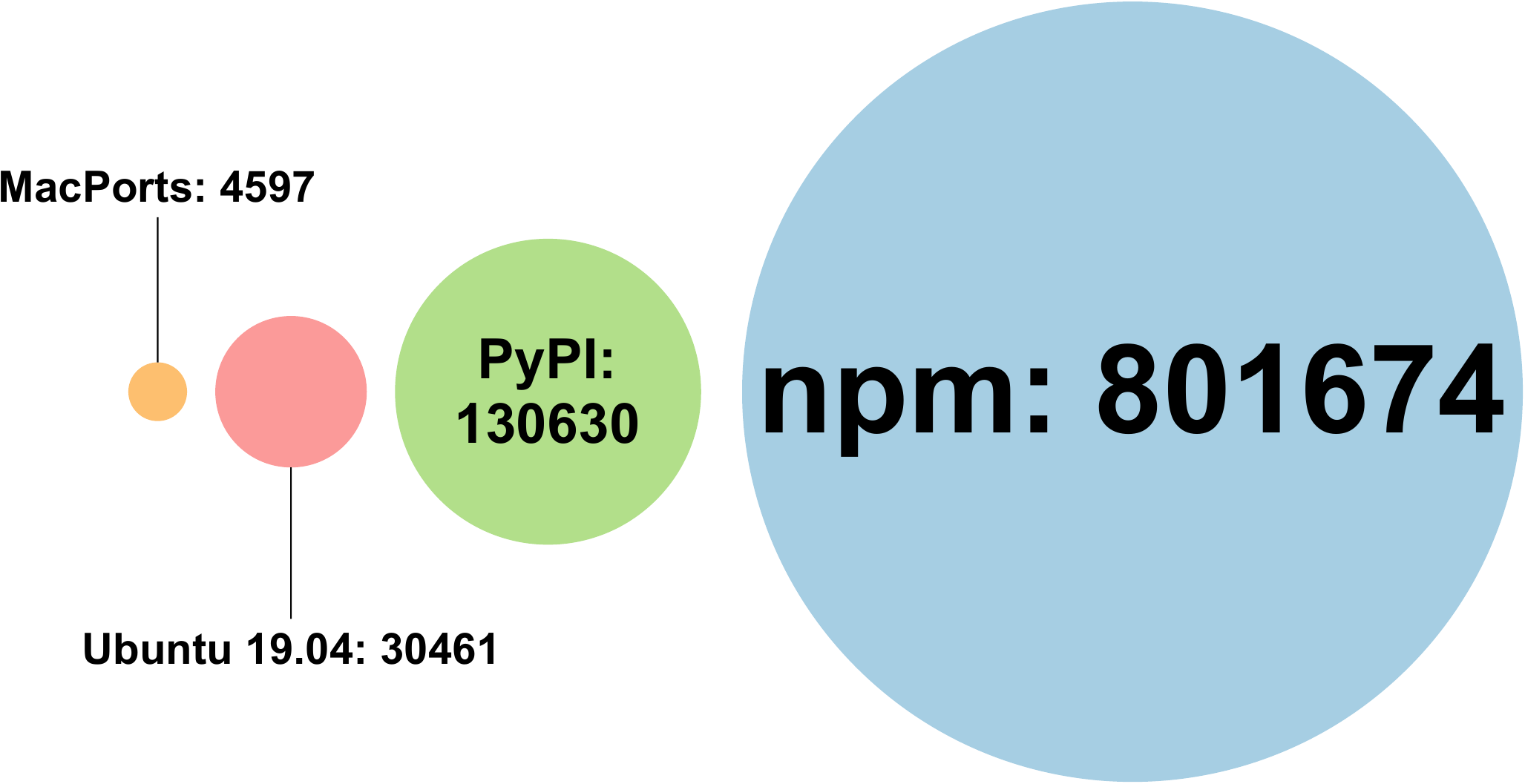}
  \caption{Ecosystem size comparison (circle area is proportional to number of packages in each ecosystem)}
  \label{fig:sizes}
  \vspace{-0.15in}
\end{figure}

\section{Analysis of Ecosystems}
\label{sec:analysis}

In this section, we elaborate upon our characterization study described in Section~\ref{sec:attacks}. We discuss the various factors that contribute to the risk and impact of attacks on LBEs in detail. We then analyze one class of attacks which has proven particularly popular, despite the apparent ease of detection: that of typosquatting.

\subsection{Methodology}
\begin{figure*}
  \begin{subfigure}{0.40\textwidth}
    \centering
    \includegraphics[width=\textwidth]{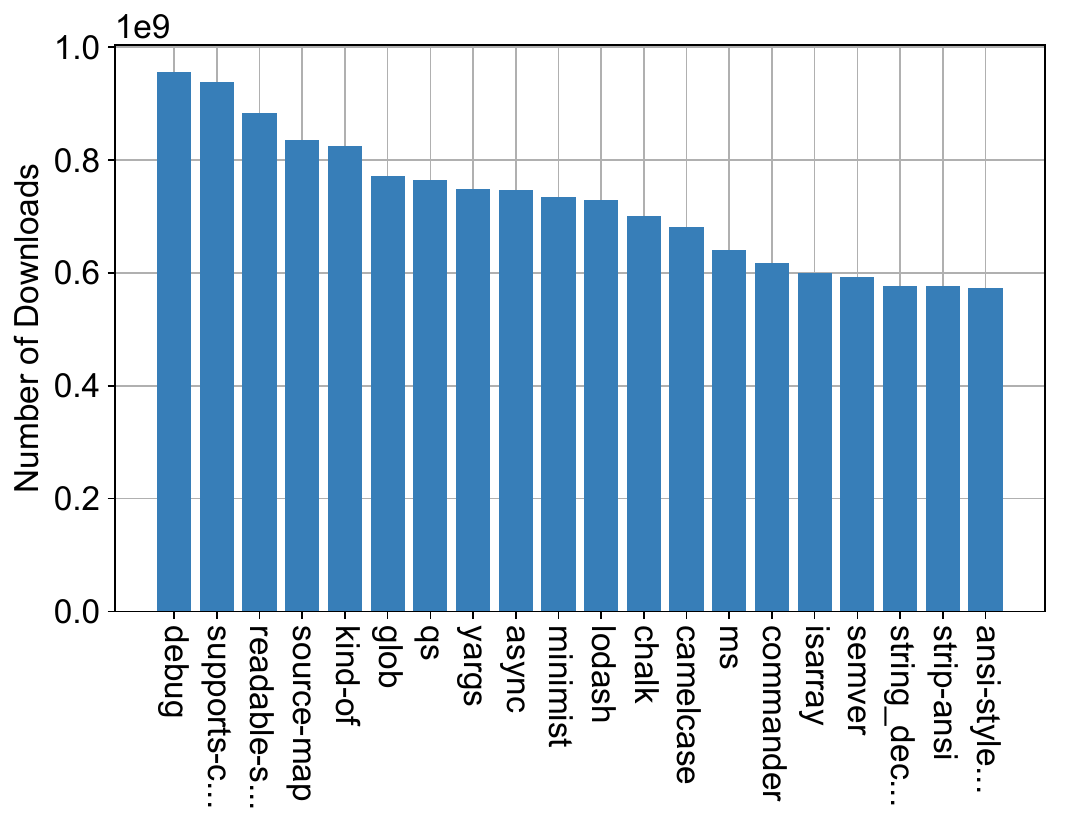}
    \vspace{-0.1in}
    \caption{npm}
  \end{subfigure}
  \hspace{2em}
  \begin{subfigure}{0.40\textwidth}
    \centering
    \includegraphics[width=\textwidth]{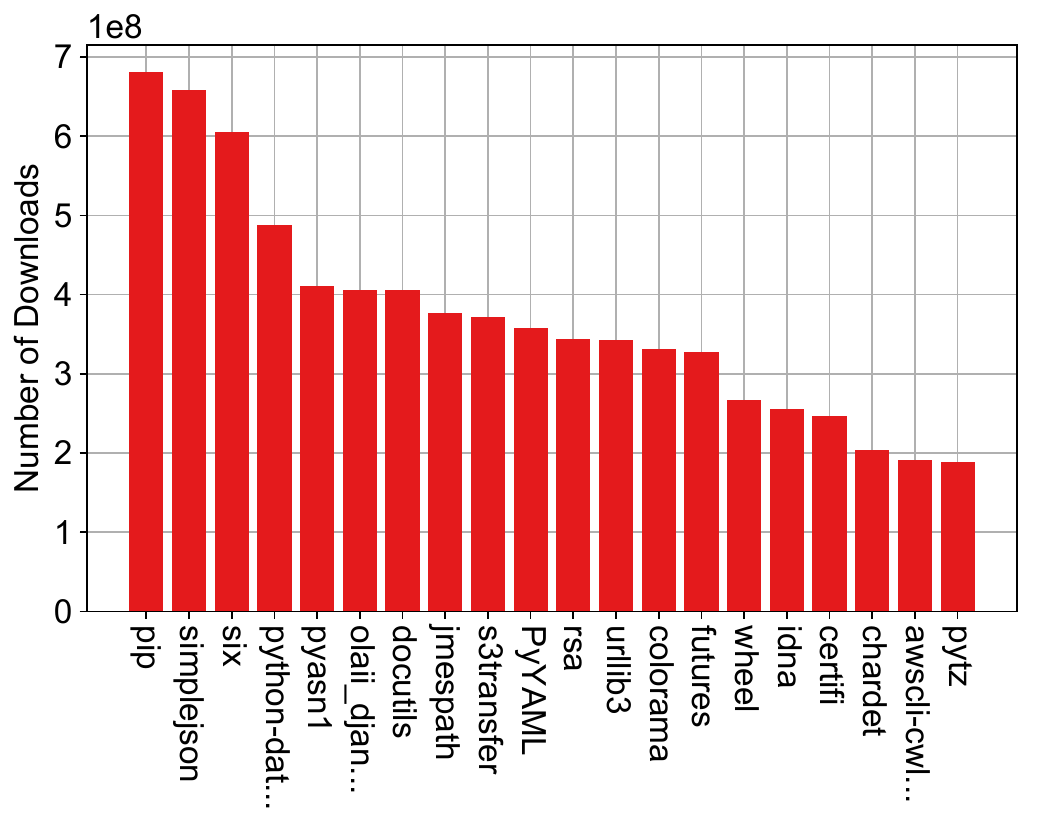}
    \vspace{-0.1in}
    \caption{PyPI}
  \end{subfigure}
  \vspace{-0.1in}
  \caption{Packages with highest download counts for each ecosystem}
  \label{fig:packagecount}
  \vspace{-0.1in}
\end{figure*}

Our analysis is based on package metadata scraped from publicly available LBEs:
npm offers a web-based API~\cite{npm_api} to retrieve metadata for the entire package dependency graph. %
It includes the number of dependencies of each package version, the author, maintainer information and other package-related metadata.
Per-package download counts are available through a different npm API endpoint~\cite{npm_download_counts_api}, which 
we used for our npm popularity findings (Figure~\ref{fig:packagecount}). 
Unlike npm, PyPI does not expose a dedicated API of dependency metadata.
Thus, we resorted to downloading each package and examining its \texttt{setup.py} script, which explicitly lists dependencies via the \texttt{install\_requires} parameter. Although examining each package's setup file has been used in prior work~\cite{ogirardot,kgullikson88}, the dependency list itself may be modified by the setup script during execution resulting in incorrect results.
Nevertheless, we feel the approach is a reasonable approximation for counting dependencies. %
For package popularity, we fetched the download information by crawling PyPI website (as it keeps the download information on the package homepage). Our npm and PyPI data snapshots were captured between December 2018 and January 2019.

\subsection{Ecosystem Evaluation}

In Section~\ref{sec:attacks}, we identified three factors that contribute to the risk of falling 
victim to an attack on an LBE: the ecosystem scale, structure, and abandonment. We evaluate 
each of these factors in turn.

\subsubsection{Analysis of ecosystem scale}

A key aspect of language-based ecosystems is their scale, that far surpasses that of traditional package management systems. Figure~\ref{fig:sizes} provides a visual comparison between the number of packages in npm and PyPI at the time of this study, to the number of packages in MacPorts (a distribution of Unix utilities for the MacOS operating system) and a recent Ubuntu version (package counts for the latter two are based on statistics from~\cite{repology_2019}). The rate of growth is also staggering: respectively 400 and 100 packages/day for npm and PyPI.

\begin{figure}[t]
  \centering
  \includegraphics[width=0.44\textwidth]{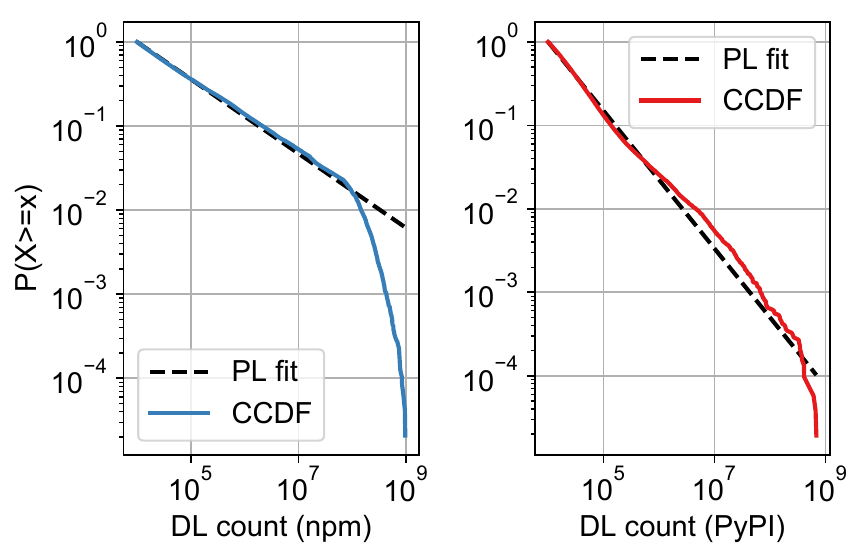}
  \vspace{-0.1in}
  \caption{Power-law distribution fit for download counts}
  \label{fig:fits}
  \vspace{-0.1in}
\end{figure}

The overall number of package downloads at the time of our snapshot was 244B for npm and 23B for PyPI. However, when looking at download counts for individual package, extreme differences emerge. For example, in npm (PyPI) the 20 most popular packages count for nearly 6\% (33\%)of the overall numbers of package downloads (note that these ecosystems include 100K to 1M packages!). A list of popular packages is presented in Figure~\ref{fig:packagecount}. We note however that there exists a long tail of packages with non-negligible ($>=$ 1000) download counts. The number of such packages is 160439 for npm, and 117524 for PyPI. We also remark---without further comment---that download counts distribution in both ecosystems appears to follow a power-law, at least for packages with high download counts (we use $10^4$ as lower bound for fitting). Figure~\ref{fig:fits} shows the CCDF of both quantities and that of the fitted distribution (for npm $\alpha=1.44$; for PyPI, $\alpha=1.83$).

In terms of scale, modern language-based ecosystems are therefore closer to mobile app stores than traditional package management systems, and are likely to suffer from similar problems. First, the scale and growth rate of these ecosystems strongly suggest that it is impossible for human maintainers to manually vet and curate the set of packages. Even automated analysis of  packages (e.g. to identify known malicious code) must be extremely efficient to keep up with the rate of growth. Second, updates to popular packages can reach a large numbers of user extremely rapidly. The combination of limited vetting and widespread distribution generates significant potential security risks due to injection of malicious packages. However, LBEs have seen little analysis, especially compared to the scrutiny of the Android and IoS mobile ecosystems (which we discuss in Section~\ref{sec:related}).

\begin{table}[!t]
  \caption{Characterization of package dependency graphs (without disconnected nodes)}
  \label{tab:graph}
  \vspace{-0.1in}
  \rowcolors{2}{gray!20}{white}
  \begin{tabular}{lcc}
    \toprule
    & {\bf npm} & {\bf PyPI}\\
    \midrule
    {\bf \#Nodes} & 577943 & 84188 \\
    {\bf Avg node outdegree} & 4.27 & 2.95 \\
    {\bf Avg dependency tree size} & 86.55 & 7.33 \\
    {\bf Avg dependency tree depth} & 4.39 & 1.71 \\
    \bottomrule
  \end{tabular}
\end{table}

\subsubsection{Analysis of ecosystems structure}
An important difference between package management systems and app stores is the \textit{interdependence} between packages. The vast majority of all packages reuse code from other packages, and/or require other packages in order to be built. As a result, in many circumstances packages may be downloaded and set up indirectly, as dependencies of other packages. Furthermore, there may be no occasion for a developer to explicitly review of the set of packages being installed a priori, e.g. if dependencies are resolved by an automated build pipeline. It is therefore crucial to understand the degree to which packages depend on each other, as the attack surface of a package effectively includes all of its dependencies.

We note that measuring package dependencies is an ill-defined problem, as package ecosystems evolve and different versions of the same package may have different sets of dependencies. In this work, we chose to use dependencies of the latest package version, as the latest version is most likely to be downloaded by users (\texttt{pip install packageName} or \texttt{npm install packageName} installs the most recent supported version by default). We also evaluated aggregating dependencies across all version of each package. As results do not change significantly, we omit them for brevity's sake.

\begin{figure}[t]
  \centering
  \includegraphics[width=0.45\textwidth]{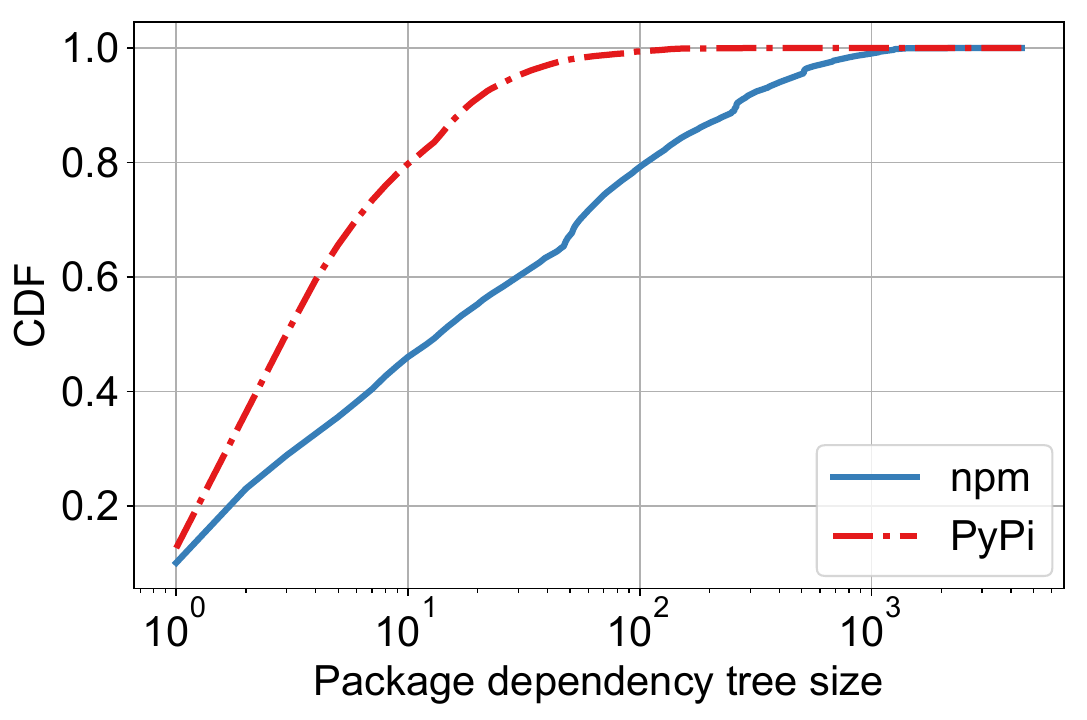}
  \vspace{-0.15in}
  \caption{CDF of dependency tree sizes for each ecosystem}
  \label{fig:dtscts}
  \vspace{-0.15in}
\end{figure}

First, we note that 28\% of npm packages and 36\% of PyPI packages do not have any dependencies or dependents. As the purpose of this section is to evaluate the potential impact of package dependencies, the statistics presented here are computed after removing these disconnected nodes from the graph. For the remaining nodes, Table~\ref{tab:graph} quantitatively characterize various properties of the package dependency graph. \textit{Node outdegree} represents the number of direct dependencies\footnote{We use the convention that a direct edge between packages $p_1$ and $p_2$ signifies that $p_1$ depends on $p_2$.}. \textit{Dependency tree size} is the number of direct and indirect dependencies (size of transitive closure), while \textit{Dependency tree depth} is the length of the longest dependency chain. Figure~\ref{fig:dtscts} presents the distribution of dependency tree sizes (transitive closures).%

Table~\ref{tab:graph} reveals that in both the ecosystems, on average, each package has more than two dependencies. Dependency trees tend to be small and shallow in PyPI, while in npm the average depth of a package dependency chain is $>$4, and the overall number of dependencies is nearly 90! This suggests a difference in development practices between ecosystems. While in PyPI developers seem to behave conservatively when incorporating external code in their packages, npm culture favors packaging and code reuse at the granularity of individual functions~\cite{haney_2016}.

Regardless of these differences, however, the analysis confirms that packages in language-based ecosystems are tightly interconnected. Figure~\ref{fig:dtscts} shows that in npm and PyPI 20\% of packages have respective more than 100 and 10 cumulative dependencies. While there is no direct negative implication for security, a large set of dependencies presents a greater attack surface on a package. Another obvious metric that can be tied to security is abandonment, which we analyze next.

\subsubsection{Analysis of package abandonment}

One of the strategies used in past incidents is to focus attack efforts against packages that are rarely or never maintained, since they are less likely to undergo scrutiny. We therefore turn our attention to such packages.

Based on past work\cite{english2007identifying}, we use lack of releases over the last 12 months as a proxy for package abandonment.  Using this metric, we calculated total abandoned packages in npm as well as in PyPI repository. We note that this metric is likely to \textit{overestimate} abandonment. Particularly in the npm ecosystem, many packages have extremely low complexity and consist of only a few lines of code; it is possible that for some of these the maintainer has simply decided that the code is perfect and does not need further updates. However, we decided to not distinguish between abandoned and completed packages, as the problem is ambiguous and ill-defined.

\begin{figure*}
  \begin{subfigure}{0.40\textwidth}
    \centering
    \includegraphics[width=\textwidth]{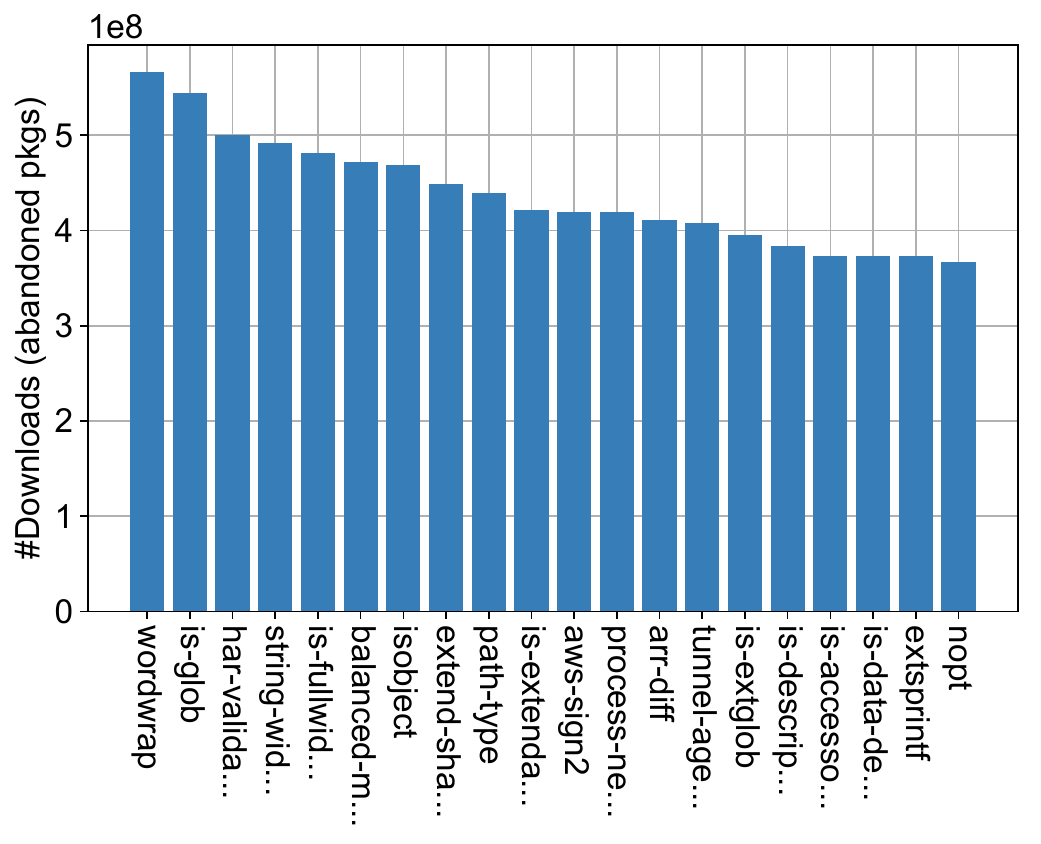}
    \caption{npm}
  \end{subfigure}
  \hspace{2em}
  \begin{subfigure}{0.40\textwidth}
    \centering
    \includegraphics[width=\textwidth]{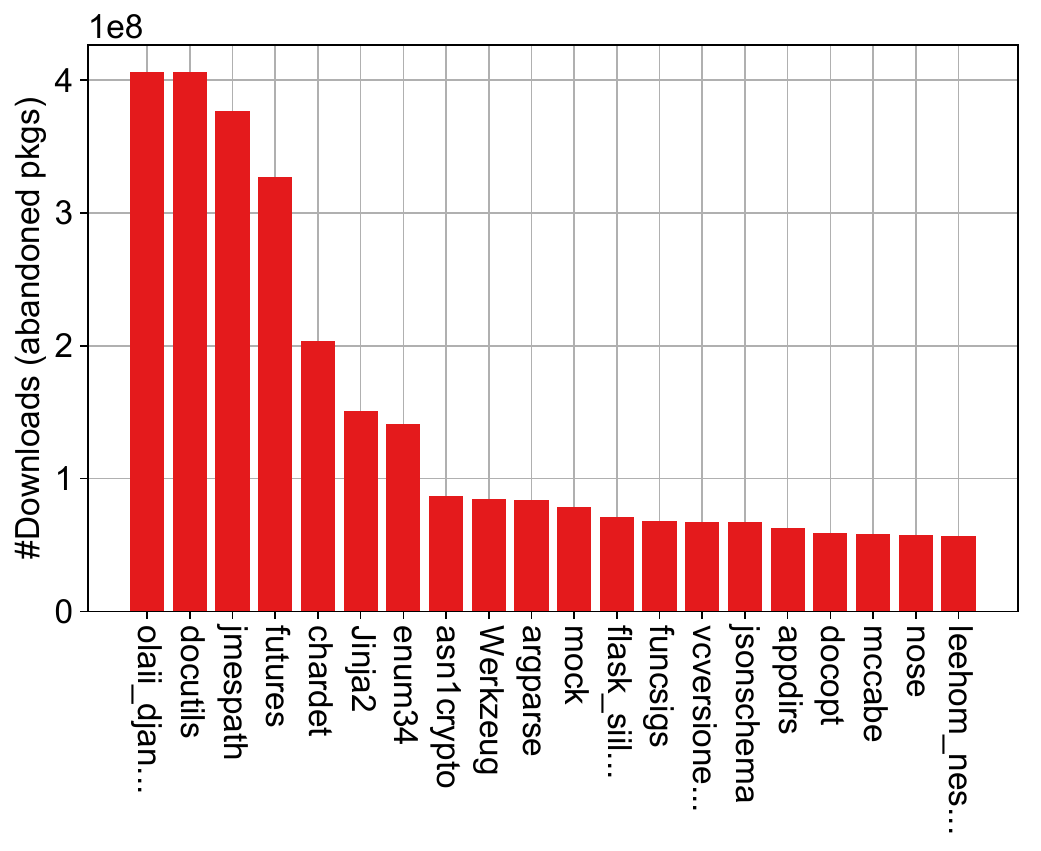}
    \caption{PyPI}
  \end{subfigure}
  \vspace{-0.1in}
  \caption{\textbf{Abandoned} Packages with highest download counts for each ecosystem}
  \label{fig:abpackagecount}
  \vspace{-0.1in}
\end{figure*}

In our analysis, we found that in npm about 496k packages (of about 801k) have been abandoned (i.e. about 61\%), while in PyPI about 74k packages (of about 130k) have been abandoned (i.e. about 57\%). Figure~\ref{fig:abpackagecount} shows abandoned packages with the highest download counts for npm and PyPI. Furthermore, while one may expect download count to be inversely correlated to probability of abandonment (i.e., that popular packages are less likely to become abandoned), we found no support for such hypothesis in the data. While this may be due to limited number of samples for high download counts, Figure~\ref{fig:abpackagecount} provides empirical evidence of abandoned packages with hundreds of thousands of download. Cumulative download count for abandoned packages approaches the billions in both ecosystems.

Overall, the fact that abandonment seems common even for highly downloaded packages is alarming because in the past attackers have successfully commandeered abandoned packages~\cite{fallingsnow_i_2018}. We discuss some possible countermeasures in Section~\ref{sec:recommendations}.

\subsection{Summary of Evaluation Results}

Our analysis determined that language-based ecosystems include a large number of packages with heavy-tailed distribution of the number of downloads, suggesting a malicious package has the potential to reach a large number of users. Furthermore, packages tend to have many dependencies, which increases their attack surface and makes it more likely for an attacker to find a viable victim package.
Combined with our finding that many highly popular packages appear abandoned by their maintainer, LBEs are fertile ground for attacks injecting malicious code into the ecosystem.
Indeed, some attacks exploiting these factors have already happened~\cite{fallingsnow_i_2018, shaun_nichols_now_2018}.

To explore these issues with more specificity, we undertook case studies for the particular issue of typosquatting in two large LBEs, npm and PyPI.
In a typosquatting attack, a developer injects a malicious package into the ecosystem with a name similar to that of another, benign package (according to the terminology of Section~\ref{sec:AttackTypes} this is a form of \textit{bait} attack). We decided to focus on this class of attacks because they have happened repeatedly in the past~\cite{npm_authors_`crossenv`_2017, goodin_devs_2017}, despite the apparent ease of detection.

\subsection{Case Study \#1: Typosquatting in npm}

In order to assess why different packages end up having similar names, we first identified all pairs of npm packages whose names differ by 1 character (i.e., string edit distance is 1). This step returned 326K candidate pairs. A quick analysis shows that most packages with very short names are part of one of such pairs (e.g., 95\% of packages with name-length = 3; this percentage decreases to 71\% for name-length = 5, and 16\% for name-length = 10). These similarities are bound to happen purely due to the size of the ecosystem: there are nearly 8000 3-letter packages in npm, and only 17576 combinations of three lowercase English letters\footnote{Names can use other symbols, however most short names do not include them.}. Therefore, attempting to identify typosquatting in short package names purely by comparing name pairs is bound to be fruitless.

One may ask whether the same approach---marking similarly-named packages as suspicious---may work at least for packages with longer names, since similarities occur less frequently. In order to answer this question, we filtered our set of similarly-named package pairs to retain only those where package names have lengths 10 or above. The filtering step returned 27K package pairs, from which we randomly selected and manually analyzed 100 pairs. %

A lone ``undecidable'' package was so marked because the distribution archive is missing from npm servers. Most of the remaining pairs (89) were labeled as ``benign''---name similarities due either to coincidence, because packages were implementing similar functionality, or to the fact that one package was derived from the other (e.g., a fork) for benign reasons. We furthermore identified nine ``suspicious'' cases.  Six consisted of empty/dummy packages named similarly to a legitimated packages, while three consisted of package pairs sharing substantial amount of codes without reasonable explanation or ties.

Finally, we marked one case as ``malicious''. This case involves the pair \texttt{agario-client} and \texttt{agario-clients}.
The latter is a near code clone of the former, which is a (now outdated) client for the agar.io browser-based game.
However, \texttt{agario-clients} includes modifications which appear to redirect users to a different game server.
Furthermore, its authors attempted to masquerade their package by manipulating package metadata in \texttt{package.json}---e.g. \texttt{\_shasum}---to mimic those of \texttt{agario-client}.
While this approach was ineffective (npm ignores such metadata as they are computed on the server side), it clearly shows the intention to make one package pass as the other.

Our brief analysis highlights a few interesting facts. First, it appears that typosquatting continues to happen in the wild, and it is likely that there are yet undiscovered cases. Second, there are no barriers - at least in the npm ecosystem - to prevent anyone from registering a package with a name close to that of a highly popular one, which may generate confusion. For example, \texttt{graphql-tools} and \texttt{graphql-tool} could be easily confused, but the former is a popular package with 500k downloads/week, while the latter appears to be a set of Angular.js programming exercises archived together. Despite not providing any usable functionality, it still managed to accrue ~600 downloads since it was uploaded. Furthermore, there are many legitimate reasons why packages may be named similarly, and most instances in which this happens are in fact benign.

Finally, since several pairs in the ``suspicious'' and ``malicious'' sets involved code cloning, one may suggest that code similarity would be a useful additional feature to distinguish false and true positives. Unfortunately, using a simple similarity metric based on file hashes revealed that nine of the benign package pairs also had significant similarities, all due to benign reasons. Overall, this analysis suggests that it may be difficult to distinguish typosquatting attacks from benign occurrences in an automated fashion.

\subsection{Case Study \#2: Import-squatting in PyPI}
Python packages present another attack vector, which we call \textit{import-squatting} and which is a variant of typosquatting. The possibility of this attack has been known to the community for some time~\cite{bs4website}. In the attack scenario, the name of the package differs from the top-level module name provided by the package. A python package provides one or more modules and it is the module name that is used to load the module through the \texttt{import} statement. It is typical for the package name to be the same as the name of the module it provides. However, this equivalence is not enforced. To consider a real-world example, the \texttt{beautifulsoup4} package, which is one of the most popular packages on PyPI, provides the module \texttt{bs4}. It is not surprising to have an unsuspecting user attempt to install this package by specifying the name \texttt{bs4} rather than \texttt{beautifulsoup4}. The risk of confusion is high enough that the authors of \texttt{beautifulsoup4} maintain a separate dummy \texttt{bs4} package to prevent someone from exploiting the problem by adding a malicious \texttt{bs4} package~\cite{bs4website}.

PyPI does not directly provide package metadata that would indicate which modules are provided by a package. We therefore downloaded the packages, and inspected their file organization to obtain information about module names. %
Because the setup script of the package can create modules during package installation and modules can be created (by other modules) during code execution, it is possible that our results are incomplete. However, we believe we have a reasonable approximation to identify modules in a package.

We found about 3,500 candidates of import-squatting. We manually analyzed 50 cases, but did not find any evidence of import-squatting. To cite an interesting example here, the package named \texttt{requirements-parser} (about 462k downloads) provides a top-level module named \texttt{requirements}. There also exists another package \texttt{"requirements"}, which we marked as likely benign, but clearly fits our criterion for import-squatting. We believe that import-squatting can be a serious threat and package maintainers should exercise due caution to protect their users.

\section{Recommendations}
\label{sec:recommendations}
In this section, we recommend ways to mitigate the threats faced by LBEs. Our recommendations are inspired by
the characteristics of attacks that we have witnessed in the
wild and leverage the metrics that we looked at 
in Section~\ref{sec:analysis}. 

We highlight two main categories of recommendations:
\begin{enumerate}
\item{{\em Technology to detect and avoid attacks in LBEs}. 
We recommend a software enhancement to package managers we call \textit{obscurity alerts} that warns users when they heuristically identified to be installing the wrong package.}
\item{{\em Best practices to employ in LBEs}. 
We recommend techniques that developers, users, and maintainers could use to avoid the circumstances that attackers have exploited in past attacks.}
\end{enumerate}

\subsection{Obscurity Alerts}
Although the popularity of a package can serve as an enticement for that package to be attacked, it can also serve as a means of defense against various ``squatting'' attacks. 
In particular, the package manager can detect when a user requests an obscure package with a name that is likely to be confused with a more popular package.
Under such circumstances, the package manager can warn the user that they may have made a mistaken request and may be about to fall victim to a squatting attack. 
We refer to such an warning as an \textit{obscurity alert}. 
The metrics that we introduce identify packages where squatting attacks are a high-risk and/or high-impact. As such, this information can be leveraged %
for issuing obscurity alerts.

As we noted in Section~\ref{sec:analysis}, package repositories already collect information that characterizes the risk and impact of squatting attacks: they can identify popular packages, and know when packages have similar names. 
We recommend that package manager software actively identify and caution users by requesting additional confirmation from before installing the near-match packages.
We note that our metrics can be used by package developers and package consumers to mitigate several 
of the attack scenarios of Section~\ref{sec:attacks}. We discuss the use of our metrics in each of these contexts below.
\subsubsection{Protecting package consumers}
The most obvious use of an obscurity alert is to protect users of package managers from falling victim to typosquatting-style attacks. 
In this regard, the most important metric for detecting an attack is edit-distance. 
However, as noted above, simply comparing a package name against all other package names presents a \naive picture of which packages are intended. Instead, when a user installs a package, we first check if it is an obscure package with a name that is close to a popular one. 

\subsubsection{Protecting package developers}
The primary way in which obscurity alerts aid package developers is the same way in which they aid application developers: by 
interposing on the direct inclusion of a package in a project 
and alerting the user to suspicious packages. 
We note that the purpose of our tool is not to prevent package developers from intentionally including obscure packages as dependencies, nor is it to prevent package consumers from relying on such packages.
We simply observe that the use of obscure packages is rare by definition and thus should be treated with heightened alertness.
As such, should an obscurity alert be raised, our tool simply asks users if they are sure that they did not intend to use the more-popular package name, and allows the installation of the obscure package with user input.

Additionally, several of the attacks described in Section~\ref{sec:attacks} relied on legitimate developers missing the actions of a malicious contributor. 
These attacks were characterized by developers who were no longer frequently updating the package, allowing an adversary (perhaps with compromised credentials) to slip a malicious update into the package under low scrutiny. The damage of such attacks is multiplied by other packages
obliviously pulling in those changes as part of an update.
Thus, we also recommend obscurity alerts be issued when pulling package updates to packages that have been abandoned (under our metric, a package is considered abandoned if it has not been updated in the last 12 months).

We emphasize that obscurity alerts are warnings only: a key finding of our work is that automatically detecting malicious packages is very difficult, and only the consumer of a package can tell if she made a typo during the request. However, this class of attacks relies on a lack of attention on the part of package consumers. If they can be alerted in the few risky cases where a typosquatting attack may happen, the threat vector becomes significantly reduced. 

\subsection{Best Practices}
Many of the attacks that we have encountered upon package 
managers could be avoided with more rigorous policies and 
greater scrutiny of package manager ecosystems. While crafting
complete audit policies is beyond the scope of this work, we
consider some general guidelines that may be helpful in 
mitigating attacks. 

\myparagraph{For package repository maintainers}
As noted above, greater scrutiny of package uploads could likely have prevented many of the known attacks on ecosystems.
Unfortunately, popular package repositories may contain hundreds of thousands of packages, so manual analysis of every package is impractical for repository maintainers.
Nevertheless, we believe that repositories should collect and report robust statistics about the use of packages such that individual package developers can leverage statistical data to identify likely mistakes. 
We also note that the more possible aliases a package uses, the greater the likelihood of confusion.
As such, we recommend that in cases where the package manager allows imports under a different name as the package itself, the difference is reported in the management tool.
Downloading a package with an import already taken by a different package may indicate confusion, and package maintainers should consider alerting users to be cautious in such cases.

\myparagraph{For package developers}
Ultimately, package developers will necessarily be responsible 
for much of the security of their package.
Broadly, we recommend that developers maintain good security
hygiene with regards to their credentials, avoid password reuse,
and update credentials frequently. We also note that when 
choosing packages to include as dependencies, package developers
take on the role of package users, and thus should follow the recommendations below.

\myparagraph{For package users}
We recommend that all developers leverage the 
metrics that we propose when considering inclusion of a package
into a project. Furthermore, when a package passes a threshold
such that it is deemed suspicious, developers should consider
checking the update, reading the changelog if present, and
determining whether the package appears to have recently changed
ownership. Such practices would have flagged many of the 
attacks seen in the past and are likely to be a good 
indicator of malicious influence in the future. 
\section{Related work}
\label{sec:related}

Literature presents many analyses of software ecosystems; however, most
do not focus on security-related aspects. Examples
include~\cite{german2013evolution, raemaekers2013maven, wittern2016look}. While
these works present useful information for understanding these complex objects,
they do not consider the issue of attackers distributing malicious software by
exploiting flaws in an ecosystem's structure.

Probably the earliest analysis of the security of a package ecosystem is Cappos et al.'s analysis of Linux and FreeBSD package managers~\cite{cappos_look_2008}. This paper predates some language ecosystems such as npm, and it is concerned with issues in the package distribution system and supporting applications rather than the ecosystem itself. Also, Athalye et al~\cite{anish_athalye_package_nodate} analyzed the security of cryptographic operations performed by package managers, such as ensuring integrity, authenticity and transport-level security. Furthermore, recent work by Pfretzschner and ben Othmane~\cite{Pfretzschner:2017:IDA:3098954.3120928} performed an analysis of possible attack techniques that a malicious package, injected as dependency, may use to attack an application. None of these works reviewed actual incidents or performed any measurement on the entity of the problem.

The work most closely related to ours is perhaps Hejderup's master thesis~\cite{hejderup_dependencies_2015}. This work quantifies the presence of vulnerable packages within the npm repository, and the extent to which other packages depend - directly or indirectly - on them. This analysis is relevant to ours, as it establishes useful practices for quantitative analysis of a package dependency graph. However, our scope is clearly different from this work---we consider attacks that are inherent to LBEs rather than those arising from software vulnerabilities.

A related line of work is on the study of application
ecosystems, most recently of mobile application markets such as
the Google Play store~\cite{viennot2014measurement,chatterjee18-ipv, wermke18,
chakradeo13}. These works are primarily concerned with
applications used by consumers, rather than application components (i.e.
packages) that are specific to the language ecosystem and are used by developers. As such, characterization of app markets (and
defenses proposed against malicious applications) are largely
orthogonal to our work. The closest work to our own is in the 
detection of \textit{cloned} applications, whereby a
lesser-known or actively malicious developer will re-package and
re-publish a better-known app. Detecting application clones has
typically been done via code similarity metrics~\cite{droidkin}
or behavior~\cite{andarwin}. In contrast, our approach is based entirely on the metadata of the entire package repository.

Other authors have looked at the more general problem of \textit{supply chain vulnerabilities}, i.e., vulnerabilities in the open-source applications on which a software package depends.
Tellnes' Master's thesis~\cite{tellnes_dependencies:_2013} investigates the effect of various classes of dependencies (including those among software components) on the reliability of a system. 
Various approaches to the containment of vulnerable dependencies are proposed, such as secure wrappers.
However, such approaches are explicitly designed for "benign" failure scenarios and unlikely to be effective against malicious dependency injections.
Cadariu et al. ~\cite{cadariu_tracking_2015} investigated vulnerable dependencies in a set of 75 production systems in the Netherlands, finding that over 70\% of these contained at least 1 vulnerable dependency. None of these studies considered maliciously-injected dependencies. Other works, such as those by Younis et al.~\cite{younis_using_2014} and Platte et al.~\cite{plate_impact_2015}, aim at reliably assessing the concrete impact of vulnerable dependencies on an application. These works are orthogonal to ours, though similar techniques could potentially be applied to determine the impact of malicious dependency injection. Finally, Kula et al.~\cite{kula_visualizing_2014} propose various plotting techniques to convey complex dependency relations. Effectively presenting this information to application developers can help avoiding dependencies on untrusted libraries.

\section{Discussion}
\label{sec:discussion}

\subsection{Limitations}
Our study is a first foray into the security issues that affect language-based ecosystems, and we make no claim for it to be exhaustive. Many of our conclusions remain qualitative, because there is just not enough evidence to support quantitative analysis. For example, data and intuition seem to support a correlation between degree of abandonment of a package and time between attack and discovery. However, many more data points would be required for a convincing statistical analysis. Furthermore, some aspects of the problem are intrinsically impossible to measure: our analysis of import-squatting is by necessity incomplete, as there are no reliable mean to identify all modules of a Python package.

Despite these limitations, we believe there is value in calling attention to ecosystem attacks, and analyzing the data which is available. These are a dangerous and novel security threat, and while high-profile incidents have occurred, analysis of the phenomenon so far has been limited. By presenting a preliminary analysis and outlining possible defensive approaches, we hope to begin a discussion within the community on how to solve these issues.

\subsection{Future Work}
We plan to explore in depth the advantages and pitfalls of various approaches to contain ecosystem attacks. For example, npm implements a \texttt{shrinkwrap} command which freezes dependencies to specific package versions. While preventing future injection attacks to succeed, however, this approach has the drawback of also locking buggy or insecure versions, not allowing them to be updated. It is unclear whether the advantages outweigh the risk of disseminating unpatchable vulnerabilities.

Even from our limited data, it is clear that letting packages run arbitrary scripts as root during installation---the model employed by OS package managers---is outdated and dangerous.
At the same time, package installation by its nature may require modifications to file system and execution of code to configure the system.
A possible solution is a access control model for package installation scripts, which would restrict installers to the minimum set of privileges necessary to accomplish their function (principle of least privilege), e.g., by restricting them to a given directory.
Further analysis is necessary to determine whether a permissions system would be effective in preventing attacks without posing unacceptable burden on package developers, especially as packages change frequently, so may change the required capabilities.

A different question is whether ecosystem attacks can be automatically detected. 
Our analysis has characterized several classes of ecosystem attacks, each working in different ways. As such, our findings indicate that there is unlikely to be a one-size-fits-all approach to ecosystem attack detection. 
Furthermore, the size of the ecosystem---and the rate at which packages are submitted~(about 100 to 400 packages daily, depending on the ecosystem)---implies that even if malicious packages were to be injected frequently, they are likely to constitute extremely rare events. 
More generally, any defense must have a minimal need for maintenance on part of the ecosystem maintainers.
Even a detector with high accuracy therefore may incur the \textit{base rate fallacy}~\cite{axelsson_base-rate_1999}---i.e., alerts are overwhelming likely to be false positives.
This conclusion is also supported by the observation, made in Section~\ref{sec:analysis} in the context of typosquatting, that benign, innocuous packages that are nearly indistinguishable from malicious ones have been generated in practice. 
All this suggests that any automated detection approach must err on the side of false negatives, to ensure that the rate at which warnings are generated remains low enough to be analyzed by human maintainers.

\section{Conclusion}
\label{sec:conclusion}

In this work, we presented a multi-faceted analysis of the security risks inherent in the structure of language-based ecosystems. Our first contribution is a taxonomy of attacks based on past incidents, which we hope will help structuring discussions and analyses of such issues. Our second contribution consists of an analysis of two representative ecosystems, the npm and PyPI ecosystems. We performed measurements of the structure of these ecosystems, and took an in-depth look at some classes of attacks, which we used as the foundation of our third contribution: a set of guidelines to contain future attacks. Overall, we found that malicious packages are hard to disambiguate from benign ones, and the scale of ecosystems prevents manual analysis by ecosystem maintainers. However, providing decision support tools to developers is likely to be effective. Indeed, when assessing whether to include an external package poses a security risks, developers are best positioned to make the decision as they are the most aware of the context.

We remark that while there has been so far a limited number of attacks, evidence suggests that these ecosystems are ripe for exploitation, and the number of incidents will only increase in the future. However, in contrast with other software ecosystems such as mobile app stores, there has been limited work on the nature of the issue and on possible solutions. By shedding light on the nature of the problem, we hope that our work will foster further research on the security of LBEs.

\bibliographystyle{ACM-Reference-Format}
\bibliography{bibliography}
\end{document}